\documentclass[12pt]{article}
\usepackage{amsfonts,amsmath,amssymb,mathrsfs,url}

\title{Comment on ``Hidden Variable Interpretation of Spontaneous Localization Theory''}
   
\author{
Roderich Tumulka\footnote{Department of Mathematics,
     Rutgers University, Hill Center, 
     110 Frelinghuysen Road, Piscataway, NJ 08854-8019, USA.
     E-mail: tumulka@math.rutgers.edu}
}
\date{August 6, 2011}

\newcommand{\be}{\begin{equation}}
\newcommand{\ee}{\end{equation}}

\renewcommand{\Im}{\mathrm{Im}}

\newcommand{\RRR}{\mathbb{R}}
\newcommand{\CCC}{\mathbb{C}}

\newcommand{\vQ}{\boldsymbol{Q}}
\newcommand{\vX}{\boldsymbol{X}}

\newcommand{\vZ}{\boldsymbol{Z}}

\newcommand{\vq}{\boldsymbol{q}}
\newcommand{\vx}{\boldsymbol{x}}

\begin{document}
\maketitle
\begin{abstract}
In his recent paper \cite{Bed}, Bedingham discussed a hybrid between a theory of spontaneous wave function collapse and Bohmian mechanics. I offer a simpler way of conveying the substance of Bedingham's paper. 

\medskip

\end{abstract}

Whereas Bedingham \cite{Bed} outlined a theory based on a continuous collapse process due to Di\'osi \cite{Dio88,Dio89}, I will describe the corresponding theory using discrete collapses, as in the original GRW theory \cite{GRW86,Bell87}. This theory, which I will call ``GRWp,'' combines elements from GRW theory and Bohmian mechanics (see, e.g., \cite{Gol01}) and is, like those, a variant of non-relativistic quantum mechanics for $N$ particles. As I will argue, GRWp is empirically equivalent to GRWf and GRWm, the GRW theories with the flash ontology and the matter density ontology (see, e.g., \cite{AGTZ06}). I am not saying, though, that GRWp should be taken seriously, or that it has advantages over either Bohmian mechanics or GRWf/GRWm.

GRWp employs a particle ontology, like Bohmian mechanics. That is, each particle has a precise position $\vQ_i(t)\in\RRR^3$ at each time $t$. Also as in Bohmian mechanics, the initial configuration is assumed to be random with distribution density $|\psi_0|^2$, and each particle moves according to Bohm's equation of motion,
\be\label{Bohm}
\frac{d\vQ_i(t)}{dt} = \frac{\hbar}{m_i} \Im \frac{\nabla_i\psi_t}{\psi_t} \bigl(\vQ_1(t),\ldots,\vQ_N(t)\bigr)\,.
\ee
However, the wave function $\psi_t:\RRR^{3N}\to\CCC$ evolves not according to the Schr\"odinger equation but according to a stochastic law similar to that of GRW. Namely, the evolution takes place as in GRW theory except that the collapse centers are not chosen randomly but are taken to be the actual particle positions---up to a random perturbation of order of GRW's collapse width $\sigma$.

In more detail, the Schr\"odinger evolution
\be
i\hbar \frac{\partial\psi}{\partial t} = -\sum_{i=1}^N \frac{\hbar^2}{2m_i} \nabla_i^2\psi + V\psi
\ee
is interrupted by discontinuous collapses that occur at random times $T_1,T_2,\ldots$ with the same distribution as in the GRW theory (viz., a collapse for particle $i$ occurs with constant rate $\lambda$). Let us consider a collapse at time $T$ affecting particle $i$; as in GRW theory, the wave function changes abruptly from $\psi_{T-}(\vq_1,\ldots,\vq_N)$ to
\be
\psi_{T+}(\vq_1,\ldots,\vq_N) = \frac{1}{C} \sqrt{g(\vq_i-\vX)}\; \psi_{T-}(\vq_1,\ldots,\vq_N)
\ee
with $C$ the normalizing constant and
\be
g(\vq) = \frac{1}{(2\pi\sigma^2)^{3/2}}e^{-\vq^2/2\sigma^2}
\ee
a 3-dimensional Gaussian function. While in GRW theory the center $\vX$ of the collapse is chosen randomly with distribution density
\be\label{vXdensity}
\rho(\vx) = \int d^3\vq_1\cdots d^3\vq_N \, \bigl|\psi_{T-}(\vq_1,\ldots,\vq_N)|^2 \, g(\vq_i-\vx)=C^2\,,
\ee
in GRWp $\vX$ is more or less the Bohmian position of particle $i$. More precisely,
\be\label{XQZ}
\vX=\vQ_i(T) + \vZ\,,
\ee
where $\vZ$ is a random 3-vector whose distribution is given by $g(\vq)$. At each collapse, a new value for $\vZ$ is chosen independently of the past. The particle positions are left unchanged when the wave function collapses; from time $T$ onwards, the particles move according to the equation of motion \eqref{Bohm} with $\psi$ the collapsed wave function. (As a consequence, the second derivative $d^2Q(t)/dt^2$ will typically be discontinuous at the time of collapse, but this is not a problem.) Then, as in GRW theory, the cycle of Schr\"odinger evolution and collapse repeats.

This completes the definition of GRWp. Now I will point out a few properties of this theory.

First, there are two sources of randomness: the initial configuration
\be
Q(0)=\bigl(\vQ_1(0),\ldots,\vQ_N(0)\bigr)
\ee
and the random perturbation $\vZ$ at each collapse. Let $T$ be the time of the first collapse and $i$ the number of the particle affected; the distribution of $Q(T)$ is $|\psi_{T-}|^2$, for the same reasons as in Bohmian mechanics: Bohm's law of motion \eqref{Bohm} transports the distribution in the same way as the Schr\"odinger equation. Now note that, as a consequence, the distribution density of $\vX$ is given by \eqref{vXdensity}, and the conditional distribution of $Q(T)$, given $\vX$, is $|\psi_{T+}|^2$:
\begin{align}
\rho\bigl(Q(T)=q\big|\vX=\vx,T=t\bigr) &= \frac{\rho\bigl(Q(t)=q,\vX=\vx\bigr)}{\rho\bigl(\vX=\vx\bigr)}\\
&=\frac{\rho\bigl(Q(t)=q\bigr)\,\rho\bigl(\vQ_i(t)+\vZ=\vx\big|Q(t)=q\bigr)}{\rho\bigl(\vX=\vx\bigr)}\\
&=\frac{|\psi_{t-}(q)|^2\,g(\vx-\vq_i)}{C^2}=|\psi_{t+}(q)|^2\,.
\end{align}

Repeating this reasoning for the subsequent collapses, we obtain three conclusions: First, at every time $t>0$ is the conditional distribution of $Q(t)$, given the times and centers of all collapses up to time $t$ (or, equivalently, given $\psi_t$), equal to $|\psi_t|^2$. Second, the joint distribution of all collapse centers and collapse times is the same as in the GRW process; thus, the entire process $\psi_t$ of GRWp has the same distribution as in the GRW theory (which justifies the name GRWp). Third and finally, GRWp is empirically equivalent to GRWf and GRWm (which are known to be empirically equivalent to each other \cite{AGTZ06}): Indeed, the flashes of GRWf are always close (viz., up to an error $\vZ$ of order $\sigma$) to the position of some particle; that is, the world line number $i$ of GRWp more or less interpolates between the flashes with label $i$ of GRWf.

A theory similar to GRWp was also considered in \cite{AGTZ06}, but with the choice $\vX=\vQ_i(t)$ instead of $\vX=\vQ_i(t)+\vZ$. Ironically, in the absence of the random perturbation $\vZ$, the theory is less transparent and harder to investigate; in particular, the relation $\rho(Q(t))=|\psi_t|^2$ no longer holds---in fact, it hardly even makes sense.

\bigskip

\noindent\textit{Acknowledgments.} 
The author is supported in part by NSF Grant SES-0957568 and by the Trustees Research Fellowship Program at Rutgers, the State University of New Jersey. 

\end{document}